\documentclass[11pt]{article}

\advance \textheight by \topskip
\usepackage{amsmath,dsfont}
\usepackage[english]{babel}

\makeatletter
\def\eqnarray{\stepcounter{equation}\let\@currentlabel=\theequation
\global\@eqnswtrue
\global\@eqcnt\z@\tabskip\@centering\let\\=\@eqncr
$$\halign to \displaywidth\bgroup\@eqnsel\hskip\@centering
  $\displaystyle\tabskip\z@{##}$&\global\@eqcnt\@ne
  \hfil$\displaystyle{\hbox{}##\hbox{}}$\hfil
  &\global\@eqcnt\tw@ $\displaystyle\tabskip\z@
  {##}$\hfil\tabskip\@centering&\llap{##}\tabskip\z@\cr}
\@addtoreset{equation}{section}
  \def\theequation{%\thesection.
  \arabic{equation}}
\makeatother

\def\cD{{{\mathcal{D}}}}

\def\II{{\mathds{1}}}
\def\IO{{0}}

\newcommand{\half}{{\scriptstyle{\frac{1}{2}}}}

\newcommand{\cJ}{{\cal J}}

\usepackage{amsmath,amssymb}
\def\beq{\begin{equation}}
\def\eeq{\end{equation}}
\def\beqa{\begin{eqnarray}}
\def\eeqa{\end{eqnarray}}
\def\barray{\begin{array}}
\def\earray{\end{array}}

\def\sc{ \scriptscriptstyle \underline }

\begin{document}
%%%%%%%%%%%%%%%%

\title{Supersymmetry  between
Jackiw-Nair and Dirac-Majorana anyons}

\author{
{\sf Peter A. Horv\'athy${}^a$}, {\sf Mikhail S.
Plyushchay${}^{b,c}$}, {\sf Mauricio Valenzuela${}^a$}\,
\footnote{e-mails:
horvathy-at-univ-tours.fr; mplyushc-at-lauca.usach.cl; valenzuela-at-lmpt.univ-tours.fr}\\
[4pt] {\small \it ${}^a$Laboratoire de Math\'ematiques et
de
Physique Th\'eorique, Universit\'e de Tours,}\\
{\small \it Parc de Grandmont,
 F-37200 Tours, France}\\
{\small \it ${}^b$Departamento de F\'{\i}sica, Universidad de
Santiago de Chile, Casilla 307, Santiago 2, Chile}\\
{\small \it ${}^c$Departamento de F\'{\i}sica Te\'orica,
At\'omica y \'Optica, Universidad de Valladolid, 47071,
Valladolid, Spain} }

\date{\today}  %\textbf{HPVLett2.tex}

\maketitle

\begin{abstract}
The  Jackiw-Nair  description of anyons combines spin-1
topologically massive fields with  the discrete series
representation of the Lorentz algebra, which has fractional
spin. In the Dirac-Majorana formulation the spin-1 part is
replaced by the spin 1/2 planar Dirac equation. The two
models are shown to belong to
 an $N=1$ supermultiplet, which carries
a super-Poincar\'e symmetry.
\end{abstract}

\vskip.5cm\noindent

\vskip.5cm
%[\texttt{arXiv:}]

%%%%%%%%%%%%%%%%%%%%%%%%%%%%%%%%%%%%%%%%%%%%%%%%%%%%%%%
%%%%%%%%%%%%%%%%%%%%%%%%%%%%%%%%%%%%%%%%%%%%%%%%%%%%%%%
%\section{Introduction}
%%%%%%%%%%%%%%%%%%%%%%%%%%%%%%%%%%%%%%%%%%%%%%%%%%%%%%%
%%%%%%%%%%%%%%%%%%%%%%%%%%%%%%%%%%%%%%%%%%%%%%%%%%%%

In the by now standard description of anyons  due to Jackiw
and Nair \cite{JN90}, the spin 1 representation carried by
the topologically massive (TM) vector system
\cite{Schonf,DJT} is combined with fractional spin. The
latter is carried by an internal space (namely the
Poincar\'e disc model of the Lobachevsky plane) \cite{Ply90}, described
by a complex coordinate $z$. A ``Jackiw-Nair'' (JN) wave
function is,
\begin{eqnarray}
    F_\mu(z,x)=\sum_n  f_n(z) F^n_\mu(x) \,\,,
    \label{Fanyon}
\end{eqnarray}
where the $f_n=c_n z^n$, $c_n=\sqrt{\Gamma(2\alpha+n)/\Gamma(2\alpha)\Gamma(n+1)}\,$, is restricted to
 the unit-disk \cite{JN90,Ply90}. The $f_n,\,n=0,1,2,...,$ span an infinite
dimensional orthonormal basis in  internal space.
$F^n_\mu(x)$ is, for each internal index $n$, a  TM wave
function. Then Jackiw and Nair propose to describe anyons
by the equations
\begin{equation}
  (P^{\mu}\mathfrak{J}^+_\mu-\beta_+m)F=0\,,
  \qquad
  \beta_+=\alpha-1,
   \label{eqfrac}
\end{equation}
where $\alpha>0$ \footnote{ The restriction to $\alpha>0$
can in fact be removed, and leads to interpolating anyons
which correspond to non-unitary representations
\cite{HPV3}.}. Here $F=(F_\mu)$ and
 $\mathfrak{J}^+_\mu$ generates the direct sum of  Lorentz algebras,
$
\mathfrak{J}^+_\mu=J^+_\mu+j_\mu,
$
 where $(J^+_\mu)_\nu{}^\lambda=i\epsilon_{\mu\nu}{}^\lambda$ generates the
spin $1$  representation of the Lorentz algebra,
 and $j_\mu$, carrying a
fractional spin, belongs to the discrete series of the
Lorentz algebra ($D^+_\alpha$) \cite{JN90}. $J^+_\mu$ acts
on the vector
 index of the field $F_\mu$, and  $j_\mu$ acts on its
``fractional'' part, labeled by $n$.
The  ``internal'' representation  can be
realized as,
\begin{equation}
j_0=z\partial_z+\alpha,
\quad
j_1=-\frac{1+z^2}2\partial_z-\alpha,
\quad
j_2=-i\frac{1-z^2}2\partial_z +i\alpha\,.
\label{holo}
\end{equation}
 The Lorentz Casimir
$j^\mu j_\mu=-\alpha(\alpha-1)$ is constant, so the
representation is irreducible.

Eqn. (\ref{eqfrac}) fixes only one of the Casimirs of the
planar Poincar\'e group, and  must therefore be
supplemented with subsidiary conditions. Those chosen by
Jackiw and Nair \cite{JN90} are equivalent to
\begin{eqnarray}
    P^\mu F_\mu=0\,,\quad
    \epsilon^{\mu\nu\lambda}P_\mu \, j_\nu \, F_\lambda=0\,.
    \label{supdjt}
\end{eqnarray}
 Eqns. \eqref{eqfrac} and  \eqref{supdjt} imply
the Klein-Gordon equation with mass $m$, while
\eqref{eqfrac} fixes the second Casimir operator of the
Poincar\'e group,
\begin{equation}
    \label{M+-}
    P^\mu\big(-\epsilon_{\mu\nu\lambda}x^\nu
    P^\lambda+\mathfrak{J}^+_\mu\big) =\beta_+m\, .
\end{equation}
Hence, the spin is $\beta_+=\alpha-1$.
Eqns. (\ref{eqfrac}) and (\ref{supdjt})
 imply
the equation of the TM theory,
\beq
    \mathfrak{D}_\mu{}^\nu F_\nu\equiv
    \left(-i\epsilon{}_{\mu\lambda}^{\ \ \ \nu}P^\lambda+
    m\delta_\mu^{\ \nu}\right)F_\nu=0\,.
    \label{DJT}
\eeq
Conversely, it can be shown \cite{HPV3} that
\eqref{eqfrac} and  \eqref{supdjt} together are equivalent
to imposing the TM
 equations \eqref{DJT}, augmented with the Majorana equation
\begin{equation}
(P^{\mu}j_\mu-\alpha m)F=0\,.
\label{maj}
\end{equation}
The Jackiw-Nair theory is, hence, equivalent to  the
coupled TM-Majorana system (\ref{DJT})-(\ref{maj}).
\vskip3mm

In another, slightly different approach  \cite{Ply91},  the anyon field  is described rather by a spinor,
\beq
    \psi_a(x,z)=\sum_n g_n(z) \, \psi^n_a(x)\,, \quad
      (P^{\mu}\mathfrak{J}^-_\mu-\beta_- m)\psi=0\,,
      \quad
      \beta_-=\alpha-\half\,,
   \label{pDir}
\end{equation}
where the
$
\mathfrak{J}^-_\mu=J^-_\mu+j_\mu
$ with  $(J^-_\mu)_a{}^b=-\half (\gamma_\mu)_a{}^b$
 generate the spin $1/2$ representation of the planar
 Lorentz group. Instead of the TM equation (\ref{DJT}),
 the field is required to satisfy the planar Dirac equation,
\beq
    {\cD}_a^{\ b} \,\psi_b\equiv(P_\mu \gamma^\mu-m)_a^{\ b}\psi_b=0\,.
     \label{Dirac}
\eeq Note that the Dirac (\ref{Dirac}) and Majorana
(\ref{maj}) equations  [with $\psi$ replacing $F$] imply
\begin{eqnarray}
    & (j_\mu \gamma^\mu+\beta_-) \psi=0,
    \qquad
    i\epsilon^{\mu\nu\lambda}P_\mu\,
    j_\nu \,\gamma_\lambda\psi=0
    &\label{supmd}
\end{eqnarray}
as consistency conditions, which eliminate the redundant
modes \cite{Ply91}.

The aim of this Note is to show that  the Jackiw-Nair and
Dirac-Majorana approaches are two facets of the same
supersymmetric system~: they are in fact, superpartners. To
this end, we note first that a fractional spin field can be
described, in {both} approaches, by equations of the same form,
\begin{eqnarray}
    {D}^\pm \psi^\pm=0,\qquad (P^{\mu}j_\mu-
    \alpha m)\psi^\pm=0,\quad \hbox{with spin }\;
    \left\{\barray{ll}
    \beta_-=\alpha-\frac{1}{2}
    \\[4pt]
    \beta_+=
    \alpha-1\,
    \earray\right.,
    \label{DMJN}
\end{eqnarray}
where  $D^+=\mathfrak{D}$  and $D^-={\cD}$ are the
operators in TM and the Dirac equations, \eqref{DJT} and
\eqref{Dirac}, respectively, and we put $\psi^-=\psi$  and
$\psi^+=F$. We note for further reference that, in both
frameworks, the posited first-order equations imply that
the field satisfies the Klein-Gordon equation.

 The (fractional) spins of the fields
$\psi^-$ and $\psi^+$ are shifted by $\beta_- -
\beta_+=1/2$, and have the same masses. They can therefore
be unified into a supermultiplet along the same lines as
done recently for the TM and  Dirac fields
\cite{HPV3,HPV4}.
 We posit
\begin{equation}
    (P_{\mu}\cJ^\mu-\hat{\alpha} m)\Psi=0,
    \qquad
    (P^{\mu}j_\mu-\alpha m)\Psi=0,
    \label{fracSM}
\end{equation}
where  $\Psi$ is formed by putting together
 the Dirac-Majorana and Jackiw-Nair \eqref{Fanyon} fields,
$
\Psi=\left(
\begin{array}{c}
\psi^-
\\
\psi^+
\end{array}\right)
$, and
$\hat{\alpha}$ is
$ \alpha_-=-\frac{1}{2}\,$ for $\psi=\left(
\begin{array}{l}
    \psi^-    \\
    0
\end{array}\right)$,
 and
$\alpha_+=-1\,$, for $\psi=\left(
\begin{array}{l}
0
\\
\psi^+
\end{array}\right).$
The first equation in (\ref{fracSM}) is the supersymmetric
equation which unifies the  Dirac and TM equations
\cite{HPV4}, and is supplemented by the Majorana equation.

The total spin operator, $ \hat{\beta}=
\hat{\alpha}+\alpha= {\rm
diag}\big(\beta_-\II_2,\beta_+\II_3\big), $ takes the value
$\beta_\pm$ on the subspaces spanned  by the
Dirac-Majorana and the Jackiw-Nair field, respectively.

Lorentz transformations are generated by
$\mathfrak{M}_\mu=-\epsilon_{\mu\nu\lambda}x^\nu
P^\lambda+\mathfrak{J}_\mu$ where
$\mathfrak{J}_\mu=diag(\mathfrak{J}_\mu^-,\,\mathfrak{J}_\mu^+)$
is block-diagonal with irreducible components acting on the
Dirac-Majorana and Jackiw-Nair components of the
supermultiplet. Augmented with translations, $P_\mu$,
yields the Poincar\'e algebra,
\begin{equation}
     [P_\mu,P_\nu]=0\,,\qquad
    [\mathfrak{M}_\mu, P_\nu]=-i\epsilon_{\mu\nu\lambda}P^\lambda\,,
    \qquad
    [\mathfrak{M}_\mu,\mathfrak{M}_\nu]=-i\epsilon_{\mu\nu\lambda}\mathfrak{M}^\lambda.
    \label{Poinc}
\end{equation}

The Lorentz algebra generated by $\cJ_\mu$  can be extended  into the
\emph{superalgebra} $\mathfrak{osp}(1|2)$ by adding the
off-diagonal matrices \footnote{Underlined capitals denote
$\mathfrak{osp}(1|2)$ spinors. }
\begin{equation}
    L_{\sc A}=\sqrt{2}
    \left(
    \begin{array}{cc}
    \IO
     & Q_{{\sc A} \,a}{}^\mu\  \\
 Q_{{\sc A} \,\mu}{}^a &  \IO
    \end{array}\right),
    \quad
     Q_{{\sc 1} a}{}^\mu=\left(
    \begin{array}{ccc}
     0 & 1 & i \\
     1 & 0 & 0
    \end{array}
    \right),\quad
    Q_{{\sc 2} a}{}^\mu=\left(
\begin{array}{ccc}
     1 & 0 & 0 \\
     0 & 1 & -i
    \end{array}\right).
     \label{5a+a-}
\end{equation}

 The operators $L_{\sc A}$ form a Lorentz spinor,
 $\big[\mathfrak{J}_\mu,L_{\sc A}\big]=
 \half\big(\gamma_\mu\big)_{\sc A}^{\ \sc B}L_{\sc B}$
 and interchange $\psi$ and $F$ (where we returned to our
original notations).

They do not preserve the physical states defined as
solutions of the Dirac and TM equations, respectively.
Consider instead the supercharges
\begin{equation}\label{Q}
     \mathcal{Q}_{\sc A}=\frac{1}{2\sqrt{m}}
     (P_\mu \gamma^\mu- R m)_{\sc A}{}^{\sc B}\, L_{\sc B},
     \end{equation}
where $R={\rm diag}\big(-\II_2,\II_3\big)$ is the
reflection operator, $\{R,L_{\sc A}\}=0$, which transform a
two-component Dirac field  into  a three-component TM field
$F'$ and conversely. Explicitly,
 \beqa
    \mathcal{Q}_{\sc A}\Psi=
    \left(\begin{array}{c} \psi'_a
    \\[6pt]
    F'_\mu
    \end{array}
    \right)
    =\left(
    \begin{array}{c}
    \sum_n f_n(z) \, \mathcal{Q}_{{\sc A} a}{}^\mu
    F^n_\mu(x)
    \\[6pt]
    \sum_n g_n(z) \mathcal{Q}_{{\sc A} \mu}{}^a
    \psi^n_a(x)
    \earray\right).
    \label{susytr1}
\eeqa
A general SUSY transformation is a linear combination of the
$\mathcal{Q}_{\sc A}$s, $\mathcal{Q}=\zeta^{\sc A}\mathcal{Q}_{\sc A}$.
Moreover,
\begin{eqnarray}
    \mathcal{D}_a{}^b \,\psi'_b&=&\zeta^{\sc A}\left(
    \mathcal{Q}_{{\sc A} a}{}^\mu \mathfrak{D}_\mu{}^\nu
    F_\nu +\frac{1}{2\sqrt{m}}
    Q_{{\sc A} a}{}^\mu (P^2+m^2) F_\mu\right) \,,
    \label{Dpsi'}
    \label{susytr2a}
    \\[4pt]
    \mathfrak{D}_\mu{}^\nu F'_\nu&=&\zeta^{\sc A}\left(-
    \frac{1}{2} \mathcal{Q}_{{\sc A} \mu}{}^a
    \mathcal{D}_a{}^b \psi_b - \frac{1}{2\sqrt{m}}
    Q_{{\sc A} \mu}{}^a  (P^2+m^2) \psi_a\right)\,,
    \label{DF'}
    \label{susytr2b}
\end{eqnarray}
showing that
$\psi'_b$ satisfies the Dirac equation if and only if $
F'_\nu$ satisfies the TM equation.

Now the Majorana equations  are intertwined by the
SUSY transformation, \beqa
    (P^{\mu}j_\mu-\alpha m) \psi'_a=\zeta^{\sc A}\left(\mathcal{Q}_{{\sc A} a}{}^\mu
    (P^{\nu}j_\nu-\alpha m) F_\mu\right)=0,
    \\[8pt]
    (P^{\nu}j_\nu-\alpha m) F'_\mu=\zeta^{\sc A}\left(
    \mathcal{Q}_{{\sc A} \mu}{}^a (P^{\nu}j_\nu-\alpha m)\psi_a\right)=0\,,
\eeqa allowing us to conclude that the $\mathcal{Q}_{\sc
A}$ in (\ref{Q}) generate indeed a supersymmetry
transformation betwen the two, DM and JN, sectors.

Completing the Poincar\'e algebra (\ref{Poinc}) by the
supercharges (\ref{Q}),
\begin{equation}
    [P_\mu,\mathcal{Q}_{\sc A} ]=0 \,,
    \qquad [\mathfrak{M}_\mu,\mathcal{Q}_{\sc A}]=
    -\frac{1}{2}(\gamma_\mu)_{\sc A}{}^{\sc B}
   \mathcal{Q}_{\sc B}\,,
    \label{SuperPoi1}
\end{equation}\vskip-7mm
\begin{eqnarray}
   \{\mathcal{Q}_{\sc A},\mathcal{Q}_{\sc B}\} &=&2(P\gamma)_{{\sc A}{\sc
    B}}
    \label{QQPab}
    +\displaystyle\frac{1}{2m}\left[({\cal J}\gamma)_{{\sc A}{\sc
    B}}(P^2+m^2)-
    2(P\gamma)_{{\sc A}{\sc B}}(P{\cal J}-
    \hat{\alpha}m)\right]\,,
    \label{SuperPoi3}
\end{eqnarray}
where $(P\gamma)_{{\sc A}{\sc B}}$ means
$P^\mu{\gamma_\mu}_{\sc A}^{\ {\sc C}}\,\epsilon_{{\sc C}{\sc B}}$.

On shell, the unified system carries therefore
 an $N=1$ superPoincar\'e symmetry.
The super-Casimir is $
 C=P^\mu{\cal J}_\mu-\frac{1}{16}[\mathcal{Q}_{\sc 1},\mathcal{Q}_{\sc 2}]=m(\alpha-3/4),
$
 a constant, showing that the representation is indeed irreducible.

%%%%%%%%%%%%%%%%%%%%%
%\section{Discussion}
%%%%%%%%%%%%%%%%%%%%%

We note, in conclusion, that the supersymmetry of the two,
Dirac-Majorana (DM) and Jackiw-Nair (JN) types of anyons proved here
explicitly had to be expected from that of the
``carrying'' spin $1/2$ and spin $1$ spaces \cite{Schonf,HPV3}.

%%%%%%%%%%%%%%%%%%%%%%%%%%%%%%%%%%%%%%%%%%%%%%%%%%%%%%%%%%%%
%%%%%%%%%%%%%%%%%%%%%%%%%%%%%%%%%%%%%%%%%%%%%%%%%%%%%%%%%%%%

\vskip 0.4cm\noindent {\bf Acknowledgements}.
 Partial support by the FONDECYT (Chile)
under the grant 1095027 and by DICYT (USACH), and by
Spanish Ministerio de Educaci\'on  under Project
SAB2009-0181 (sabbatical grant of MSP), is acknowledged. MV
has been supported by CNRS postdoctoral grant (contract
number 87366).

%%%%%%%%%%%%%%%%%%%%%%%%%%%%%%%%%%%%%%%%%%%%%%%%%%%%%%%%%%%%
%%%%%%%%%%%%%%%%%%%%%%%%%%%%%%%%%%%%%%%%%%%%%%%%%%%%%%%%%%%%

\end{document}